\documentclass[conference]{IEEEtran}
\ifCLASSINFOpdf
  \usepackage[pdftex]{graphicx}
  \graphicspath{{Figures/}}
 \DeclareGraphicsExtensions{.pdf,.jpeg,.png}
\else
\fi
\usepackage[cmex10]{amsmath}
\usepackage{array}
\usepackage[caption=false,font=footnotesize]{subfig}
\usepackage{amssymb}
\usepackage{amsmath}
\usepackage{soul}
\usepackage{algorithm}
\usepackage{algorithmic}
\usepackage{float}
\newfloat{algorithm}{t}{lop}

\usepackage{stfloats}
\begin{document}
%
\title{A Dynamic Embedding Algorithm for Wireless Network Virtualization}

\author{\IEEEauthorblockN{}
Jonathan van de Belt, Hamed Ahmadi, Linda E. Doyle
\IEEEauthorblockA{CTVR - The Telecommunications Research Centre
University of Dublin, Trinity College \\
Email: \{vandebej, ahmadih, linda.doyle\} @tcd.ie}}


%


\maketitle

\begin{abstract}
Wireless network virtualization enables multiple virtual wireless networks to coexist on shared physical infrastructure. However, one of
the main challenges is the problem of assigning the physical resources to virtual networks in an efficient manner. Although some work has
been done on solving the embedding problem for wireless networks, few solutions are applicable to dynamic networks with changing traffic
patterns. In this paper we propose a dynamic greedy embedding algorithm for wireless virtualization. Virtual networks can be re-embedded
dynamically using this algorithm, enabling increased resource usage and lower rejection rates. We compare the dynamic greedy algorithm to a
static embedding algorithm and also to its dynamic version. We show that the dynamic algorithms provide increased performance to
previous methods using simulated traffic. In addition we formulate the embedding problem with multiple priority levels for the static and dynamic case.
\end{abstract}


%
\IEEEpeerreviewmaketitle

\section{Introduction}
Mobile network operators are currently experiencing large challenges with the continued growth of mobile data traffic. The volume of data traffic is expected to increase annually by 66\% for the period 2012-17 \cite{Cisco2013}. However the revenues collected by mobile network operators are not growing at the same rate, while deploying new technologies is increasingly expensive \cite{NEC2013}. The sharing of physical infrastructure resources has been a traditional means of reducing both CAPEX and OPEX costs for operators. The passive sharing of physical sites, tower masts and other support systems is well established. So too is the model of the Mobile Virtual Network Operator (MVNO). The MVNO does not own the wireless infrastructure but instead enters into a service level agreement with a mobile network operator to obtain bulk access to network services at wholesale rates, which it uses to serve an independent customer base. The service level agreements in place are simple. The MVNO bulk-buys minutes or data, based on coarse MVNO usage level predictions (for multiple months) and corresponding spare capacity predictions by the MNO.  Typically there is no differentiation between the MVNO or the MNO user on the physical network.



The word virtual in the phrase MVNO simply serves to emphasize the lack of ownership on the part of the MVNO of the physical infrastructure and the spectrum. In recent years, however, virtualization techniques have been emerging which focus on what is termed active sharing of spectrum and infrastructure \cite{Costa2013}. This allows multiple heterogeneous virtual networks to coexist concurrently on a shared physical infrastructure in a much more sophisticated and resource efficient manner than is the case for MVNOs. The virtual network operators (VNOs) of the future will be able to request resources from the physical infrastructure provider (InP) based on more specific requirements and for specified time slots rather than based on needs averaged over long periods of time. Through better use of resources there is potential to support greater numbers of virtual operators on a given infrastructure and to potentially provide differentiated experiences for users from different virtual operators.


Any virtualized system requires full isolation between virtual entities. Since the wireless links are broadcast in nature and are influenced by interference, it is first necessary to divide the wireless resources into orthogonal isolated dimensions [5]. One of the fundamental challenges in wireless network virtualization is how to assign the isolated resources to the different virtual operators in an efficient and optimal manner. This is also known as the virtual embedding problem and is the focus of this paper.

Currently there are a number of wireless network virtualization solutions in existence which deal with the embedding problem. In \cite{Zaki2011} the authors focus on virtualizing Long Term Evolution (LTE) networks. A ”Hypervisor” layer is proposed that is added on top of the physical responsible for the resource and spectrum allocation to VNOs. However, it is difficult to design a hypervisor that can perform efficient resource allocation in real time. Two other approaches are the network virtualization substrate (NVS) \cite{Kokku2012} and Cellslice algorithms \cite{Kokku2013}. These approaches are based on using slices to separate the virtual networks and enabling customization of slices to utilize resources efficiently. The authors of \cite{Yang2012} present a Karnaugh-map based approach to deal with the problem of assigning virtual network requests to a 2-dimensional substrate.
Although these approaches have made significant contributions, they do not address heterogeneous networks and it is not possible to alter the slices dynamically; in other words once a virtual operator is assigned resources, the assignment cannot be changed.




The purpose of this paper is to explore how the virtual embedding process can be made more efficient and dynamic. The term dynamic refers to embedding algorithms that can re-embed existing virtual networks to achieve more efficient resource allocation and satisfy additional virtual network requests \cite{Fischer2013}. We present a dynamic greedy algorithm based on the combination of virtual networks that maximise resource usage and profit, and also adapt an existing approach to be dynamic. Using this dynamic approach, the virtual networks can be re-embedded at every timeslot, to allow for more efficient resource allocation and fulfillment of the needs of virtual network operators. Both the physical infrastructure provider and the VNOs benefit from this, as the costs decrease for the VNOs and the revenue increases for the infrastructure provider.

The virtualization model and the concept of dynamic embedding are described in section \ref{sect:model}. This section of the paper also makes one of the contributions in providing a formulation for the embedding problem with priority levels. Section \ref{sect:implementation} describes the second contribution of the paper; the greedy dynamic embedding algorithm and an adapted algorithm based on the Karnaugh-map algorithm \cite{Yang2012}. Section \ref{sect:evaluation} provides the key results and shows how the dynamic approach significantly improves on a static embedding approach and we conclude in Section \ref{sect:conclusion}.

\section{System Model}
\label{sect:model}


Wireless virtual network embedding requires the division of substrate resources into orthagonal dimensions to prevent link interference \cite{Park2009}. For now, we assume that we are only concerned with a single base station, i.e. one geographical location so that the virtual networks do not change location when reassigned. The wireless substrate is divided into a number of frequency and time-domain resources, represented as $F$ and $T$. The total number of virtual resources available is $F \times T$.

Each virtual network requires a number of frequency and time-domain resources. Virtual network operators can make requests for these physical resources, known as virtual network requests (VNRs). VNRs have different priorities, based on the service level that the VNO wants to provide to its services/users. In exchange for providing the resources, the InP obtains revenue from the VNOs. The revenue depends on the number of resources used, the duration for which they are in use and the priority level, since requests with higher priorities are more likely to be embedded instantly, and therefore should have higher costs.

The resource allocation is performed at regular timeslots. A number of VNRs can arrive during a timeslot. These requests are stored in a buffer and at the beginning of the next timeslot, the resource allocation is performed for the VNRs in the buffer. The set of VNRs at timeslot $t$ are represented by $\mathcal{R}_t$. Each request is a set $R = (p, f, td, d)$. The priority level is represented by $p$, and the frequency and time-domain requirements are denoted by $f$ and $td$ respectively. Each VNR requires an area $A$ of $f \times td$. The duration of timeslots for which the resources are required is represented by $d$.

A VNR is successful when the required resources are allocated to the virtual network. Once a VNR is successfully embedded it will retain the resources needed until it expires. If a VNR is unsuccessful within a given timeslot, the VNR remains in the buffer and is embedded in the next timeslot if possible. When a VNR has not been successfully embedded for a certain number of timeslots, $max\_delay$, it is removed from the buffer and the request is rejected. The maximum delay can be different for every priority level.

\begin{figure}[!t]
\centering
   \subfloat[Static embedding]{%
     \includegraphics[width=.46\textwidth]{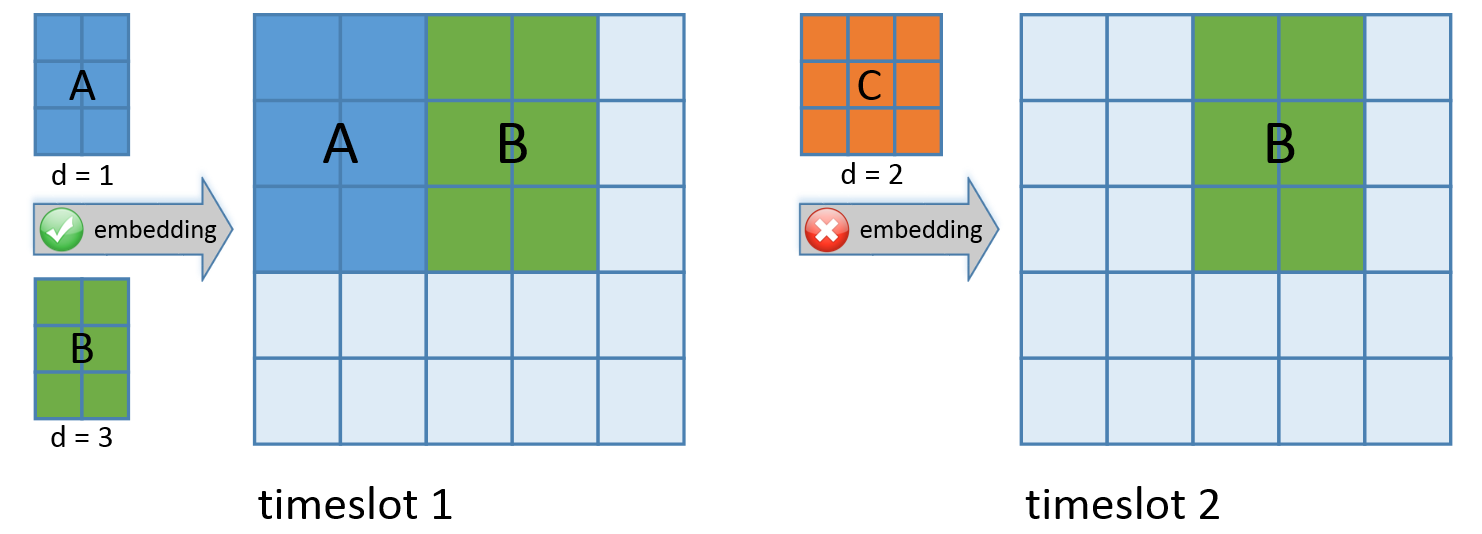}}\hfill
   \subfloat[Dynamic embedding]{%
     \includegraphics[width=.46\textwidth]{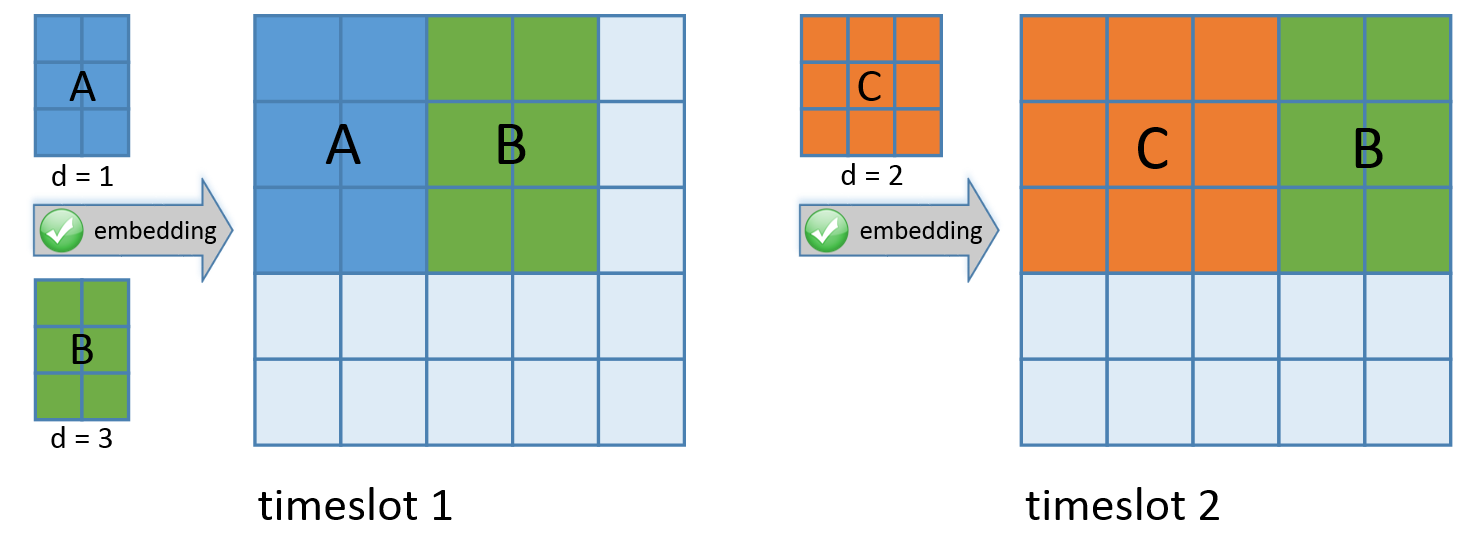}}\hfill
   \caption{Using static embedding, requests A and B can be embedded in timeslot 1, however request C is rejected in timeslot 2. When using dynamcic embedding, request C can be accepted in timeslot 2.}\label{fig:1}
\end{figure}

\subsection{Dynamic Embedding}

In the static case, successful VNR will receive the same set of resources every timeslot for the duration specified. However with this static embedding scheme it is possible that a virtual network request is rejected although there are enough resources available to satisfy it due to topology limitations.

The topological limitations of the static embedding scheme are explained in figure \ref{fig:1}a. Let us assume that the substrate has $5 \times 5$ resources available. At timeslot 1 there are two VNRs A and B with area requirements of $2 \times 3$ and durations of 1 and 3 respectively. Since there are enough free resources, both networks are embedded. At the next timeslot, virtual network A has expired and the resources are released. At the same time a VNR arrives with a resource requirement of 3 frequency and 3 time-domain resources. The static embedding method is not able to accept request C because it cannot change the assigned resources of the embedded request B.

It is possible to overcome this problem by dynamically reassigning the existing networks at every timeslot. The resources can be used much more efficiently and additional VNRs can be accepted. Going back to our example, it is clear from Figure \ref{fig:1}b that using the dynamic embedding approach the substrate resources are used in a more efficient manner.

There some considerations that must be taken into account when the virtual network embedding is performed dynamically. Dynamic embedding is more complex than static embedding in the hardware side, and has larger overheads. It is also easier to define the heuristics for static embedding. However it might be beneficial for InPs to use more expensive hardware to support dynamic embedding if the performance is better than static embedding.





\subsection{Problem Formulation}
The allocation of physical resources to virtual network requests is performed at the start of each timeslot. The new requests with the highest priority are embedded first, followed by the lower priorities. To this end, the resource allocation is performed over several stages, corresponding to each priority level.
In the static case, we must take into account the resources that are occupied and ensure that VNRs are embedded only in locations that are vacant. In contrast in the dynamic case, we must guarantee that existing virtual networks receive the necessary resources before embedding new VNRs. By introducing an additional stage for the existing networks, we can ensure that they will be embedded first, followed by the VNRs in descending priority order.

The embedding of the existing virtual networks and the new VNRs is performed in stages from 1 to the number of priority levels, $K$. At each stage, $k$, the embedding for the previous stages must be taken into account. $\mathcal{P}_k$ represents the set of virtual networks embedded in the stages previous to stage $k$. The set $\mathcal{P}_1$ is always be empty as there are no previous stages that need to be accounted for. The set of requests embedded at stage $k$ is represented as $\mathcal{S}_k$. The full embedding for a timeslot is thus found when the embedding for the last stage ($k = K$) has been completed.

The objective at each timeslot $t$ is to maximize the revenue for the infrastructure provider while guaranteeing that VNRs are embedded in order of priority. We propose a revenue function, $rev(t)$, which represents the revenue that the infrastructure provider can achieve at every time slot. The revenue collected is based on the number of resources occupied at each timeslot and their priority level. We assume that the physical infrastructure owner can earn more revenue for high priority virtual networks than for networks with lower priorities which is shown by the scalar $p_s$ .

\begin{align}
\label{eqn11}
& rev(t) = \sum_{s=1}^{|\mathcal{P}_K|+|\mathcal{S}_K|} rev_s(t) \\
\label{eqn12}
& rev_s(t) =\begin{cases}
    p_{s} Area_s, & \hspace{5mm} \text{if embedded}.\\
    0,  & \hspace{5mm} \text{otherwise}.
  \end{cases}
\end{align}

The binary variable $a_{sij}$ represents the possible embedding locations for each virtual network. The frequency and time dimensions of the substrate are $i \in \{1, 2, \cdots, F\}$  and $j = \{1, 2, \cdots, T\}$ respectively. When $a_{sij} = 1$ this means that the virtual network $s$ is using the resource block $i, j$ and when $a_{sij} = 0$ it is not using the resource block.

We also add an auxiliary binary variable $b_{sij}$ that is used to represent the starting position of the virtual network. For any particular VNR $s$, there can only be one position at which $b_{sij} = 1$.

The dynamic resource allocation problem is represented by the following optimization problem for every priority level $k$.

\begin{subequations}\label{grp}
\begin{align}
\label{eqn1}
 \max_{a_{sij} \hspace{1mm} b_{sij}} &  \sum_{s=1}^{|\mathcal{P}_k|+|\mathcal{S}_k|} \sum_{i=1}^{F}  \sum_{j=1}^{T} p_s a_{sij}, &&   \\
\label{eqn2}
  \text{S. t.} \hspace{8mm} &  \sum_{i=1}^{F}\sum_{j=1}^{T} b_{sij} = 1 && \forall s \in \mathcal{P}_k, \\
\label{eqn3}
 &  \sum_{i=1}^{F}\sum_{j=1}^{T} b_{sij} \leq 1 && \forall s \in \mathcal{S}_k, \\
\label{eqn4}
 &  \sum_{s=1}^{|\mathcal{P}_k|+|\mathcal{S}_k|} a_{sij} \leq 1 && \forall i, j,  \\
\label{eqn5}
 &  \sum_{i=1}^{F}\sum_{j=1}^{T}(a_{sij} \!- \!b_{sij}.A(s)) = 0 && \forall s \in \mathcal{P}_k, \mathcal{S}_k , \\
\label{eqn6}
 &   b_{sij}.A(s) \!\!-\!\!\!\! \sum_{p=i}^{i+f-1} \sum_{q=j}^{j+td-1}\!\!\!a_{spq} \leq 0 \! \!&& \forall s, i, j, \\
\label{eqn7}
 &  a_{sij} \in \{0, 1\}, & \\
\label{eqn8}
 &  b_{sij} \in \{0, 1\}, &
\end{align}
\end{subequations}

\noindent where $A(s)$ is the area of request $s$. \eqref{eqn2} ensures that the existing requests must be embedded, while \eqref{eqn3} shows that the requests with lower priority can be embedded. \eqref{eqn4} provides the constraint that virtual networks must not overlap. \eqref{eqn5} ensures that only one $b_{sij}$ can be selected as starting point for each request and \eqref{eqn6} provides the dimensional constraints.

In the static case we must also consider the existing substrate, $e_sij$. This matrix represents the locations of the existing networks which must be re-embedded in the same location. The optimization problem is similar to \eqref{grp} but we also have the additional constraint below to ensure that existing networks are mapped to the same resources.

\begin{align}
\label{eqn9}
& & e_{sij} - a_{sij} \leq 0 \hspace{12mm} \forall s, \forall i, \forall j
\end{align}

For both the static and the dynamic case, this problem is repeated for all stages 1 through $K$ to find the overall resource allocation.

\section{Dynamic Embedding algorithms}
\label{sect:implementation}

The embedding problem described in Section \ref{sect:model} is a more complex version of the set packing problem \cite{Crescenzi1998}, with the additional requirement of priority levels. The knapsack problem is NP-complete and thus heuristic algorithms are necessary to achieve feasible solutions.

\subsection{Karnaugh-map algorithm}
The Karnaugh-map based approach in \cite{Yang2012} is an example of a heuristic solution. In this work, heterogeneous VNRs are embedded on a two-dimensional substrate by finding Karnaugh-map like regions of vacant resources. We adapted the algorithm to suit the dynamic case. A detailed description is given below.

The aim of the Karnaugh-map algorithm is to maximize the resource occupancy of the substrate and minimize the rejection rate of VNRs. The algorithm attempts to place virtual networks on the substrate in such a manner that the highest number of additional VNRs can be accepted. The best way of achieving this is by clustering the virtual networks together, so that the vacant resources are contiguous. This allows virtual networks of varying sizes to be embedded, maximising the occupied area and minimising the rejection rate.

At each timeslot the set of VNRs that are in the buffer are sorted by priority and by decreasing area. Then the requests are embedded sequentially; the existing virtual networks are re-embedded first and the new requests follow. For each VNR, the Karnaugh map approach is used to find the set of vacant substrate regions with dimensions equal to or greater than the virtual network's $f$ and $td$. The smallest region is selected from this list as the location for embedding the VNR. Within this region it is necessary to find the best corner (of the four corners) at which to embed the virtual network.

The corner that maximizes resource clustering is found using the Embedding Density Index (EDI). The EDI measures the number of borders between free and occupied resource blocks. A high EDI index represents a substrate where the virtual networks are spread out and not clustered together, whereas a substrate with virtual networks that are grouped together will have a low EDI index. For each possible corner, the EDI is calculated by placing the new virtual network on the current substrate affects the clustering of resources. After testing the possible locations, the location with the smallest EDI is chosen as the location to embed the new network. This ensures that the network is grouped with the embedded networks as best as possible.

The duration remaining for each of the networks is updated and networks that have expired are removed.

%

\subsection{Greedy algorithm}

We developed a heuristic algorithm known as the Greedy dynamic algorithm. Rather than sorting the VNRs by area, this algorithm considers the embedding of combinations of multiple VNRs, so that a more efficient embedding can be found. Similar to the Dynamic Karnaugh-map algorithm, the input at each timeslot is the set of existing virtual networks and the VNRs in the buffer. The aim is to find the combination of virtual networks that maximizes the number of embedded virtual networks, while guaranteeing that VNRs with high priority are considered first.

As is shown in Algorithm \ref{Algorithm:1}, it is first necessary to find the set of possible combinations of VNRs. For each priority level, the set of combinations of higher priority VNRs must be included, but since the higher priority requests are allocated resources first, the combination that maximises the higher priority requests must be selected (referred to as $previous\_combination$). $previous\_combination$ is known already, and thus it is only necessary to find the set of combinations of VNRs with the current priority level.

Similarly, the combination of existing virtual networks is known at the start of each embedding. This combination can be included at the highest priority level by simply substituting $previous\_combination$ with the  combination of existing networks.


The set of combinations are sorted by decreasing total number of resources used. Although we know that it is possible to embed the combinations based on resource constraints, we cannot be sure that it is possible to embed the combinations based on the dimensional constraints of each VNR. To overcome this, we test the embedding of combinations sequentially until an allowed embedding is found. The algorithm used to perform the embedding is the Dynamic Karnaugh-map; however any embedding algorithm could be used.

\renewcommand{\algorithmicrequire}{\textbf{Input:}}
\renewcommand{\algorithmicensure}{\textbf{Output:}}

\begin{algorithm}
\caption{Dynamic Greedy Embedding Algorithm}
\label{Algorithm:1}
\begin{algorithmic}[1]
 \REQUIRE $\mathcal{R}_t$ and $previous\_combination$
 \FORALL{priority levels}
 \STATE{possible combinations, $P = \{\}$}
 \STATE{$S$ =  subset of $\mathcal{R}_t$ with current priority}
 \STATE{find all combinations, $C = \binom{|S|}{n}$ for $n=1:|S|$}
 \FORALL{$i=1:|C|$}
 \STATE{$T = \{previous\_combination\} \cup \{C(i)\}$}
 \IF{area of VNRs in $T <= Total\_Area$}
 \STATE{$P =[P; T]$}
 \ENDIF
 \ENDFOR
 \STATE{sort $P$ by area of combinations}
 \WHILE{no embedding found}
    \STATE{Dynamic Embedding for current combination in $P$}
 \ENDWHILE
 \STATE{$previous\_combination = $current combination in $P$}
  \ENDFOR

\end{algorithmic}
\end{algorithm}

\section{Performance Evaluation}
\label{sect:evaluation}
In this section we measure the performance of the static and dynamic Karnaugh-map algorithms. The metrics used to evaluate the embedding algorithms are discussed and we examine how the algorithms compare to the solutions found through MIP-optimization.

\subsection{Evaluation metrics }

The first metric we use to evaluate algorithm performance is the revenue that the physical infrastructure can expect to achieve from the virtual network operators.

The second metric that is used is the rejection rate of virtual network requests, $r(t)$, which reflects the level of service that the VNOs receive from the InP. The rejection rate per timeslot depends on the number of requests that were rejected, $s_{rej}$, over the total number of requests. The area, $A_s$, and duration, $d_s$, of each request are taken into account.

\begin{align}
\label{eqn10}
r(t) = \sum_{s=1}^{\mathcal{R}_t} \frac{ A_{s, rej}d_{s, rej}}{{A_s}{d_s}}
\end{align}

We do not compare the complexity of the algorithms since the optimal implementation of each algorithm is not known and so a fair comparison is not possible.

\subsection{Simulation setup}
We assign three priority levels for virtual network requests; the highest for real-time services such as VOIP, the second for high data rate services such as video streaming and the lowest priority for best effort services such as messaging and downloading.

The arrival of requests is modelled as a Poisson process with average rate $\lambda$ virtual networks arriving per timeslot. The duration of each virtual network follows an exponential distribution with an average lifespan of $\mu$ timeslots. The default values are $\lambda = 3$ and $\mu = 10$. The priority level of each new VNR is distributed uniformly as $U(1, 3)$. The maximum delay, $max\_{delay}$, is set to 1 timeslot for high priority requests, 2 timeslots for medium priority requests, and 3 timeslots for low priority requests. The dimensions of the substrate are set as $F = 12$ and $T = 12$.

The area of each request is initially modelled as a uniform distribution with $U(1, 3)$ for the frequency dimension and $U(1, 3)$ for the time dimension. We run the simulation for $1000$ iterations. The function $p_s$ which represents the relative cost for different priority levels is modelled as follows $p_s = \alpha$ for the highest priority, $\beta$ for the medium priority and $\gamma$ for the lowest priority. We set $\alpha=0.5$, $\beta=0.3$ and $\gamma=0.2$.

\subsection{Results and discussion}

\begin{figure}[!t]
\centering
  \subfloat[Revenue]{%
    \includegraphics[width=.46\textwidth]{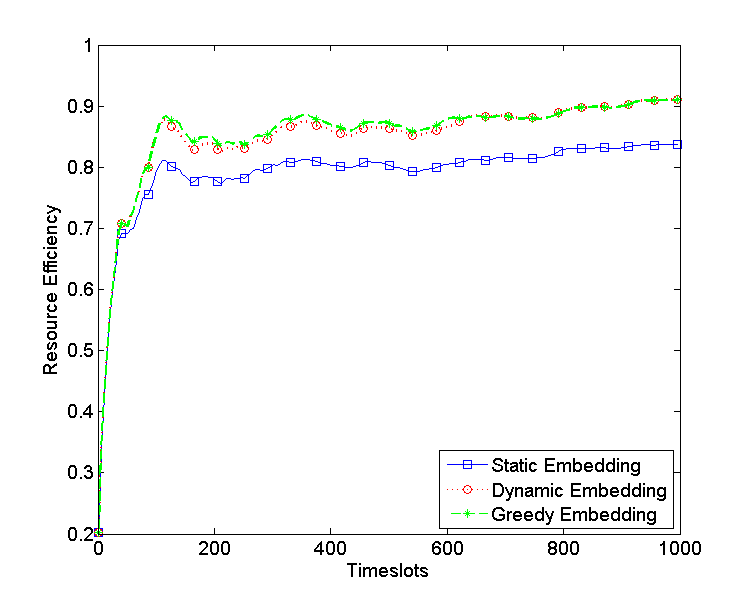}}\hfill
  \subfloat[Rejection Rate]{%
    \includegraphics[width=.46\textwidth]{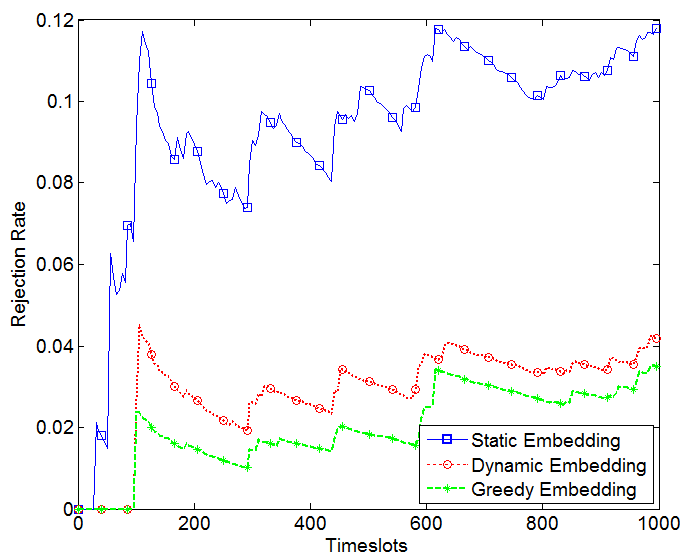}}\hfill
  \caption{The dynamic embedding algorithms provides an increase in revenue for the InP and a large decrease in the rejection rate of virtual network requests.}\label{fig:3}
\end{figure}

\begin{figure}[!t]
\centering
 \subfloat[Revenue]{%
    \includegraphics[width=.46\textwidth]{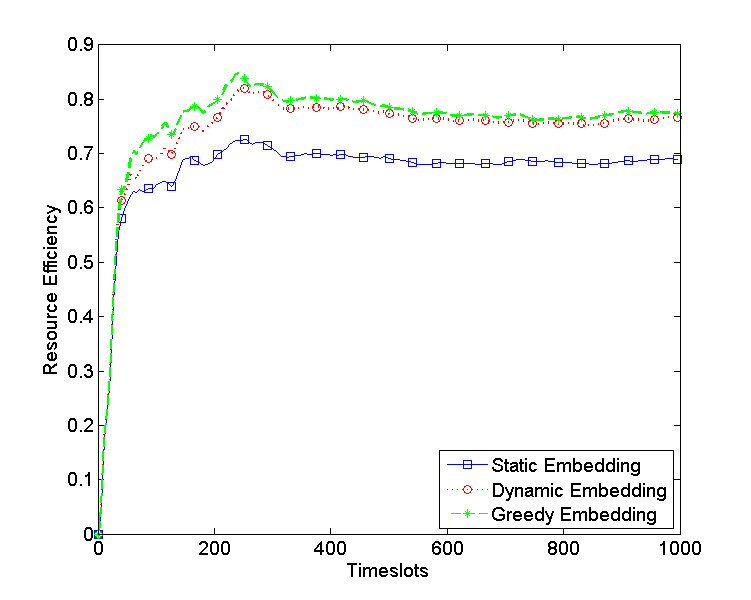}}\hfill
  \subfloat[Reject Rate]{%
    \includegraphics[width=.46\textwidth]{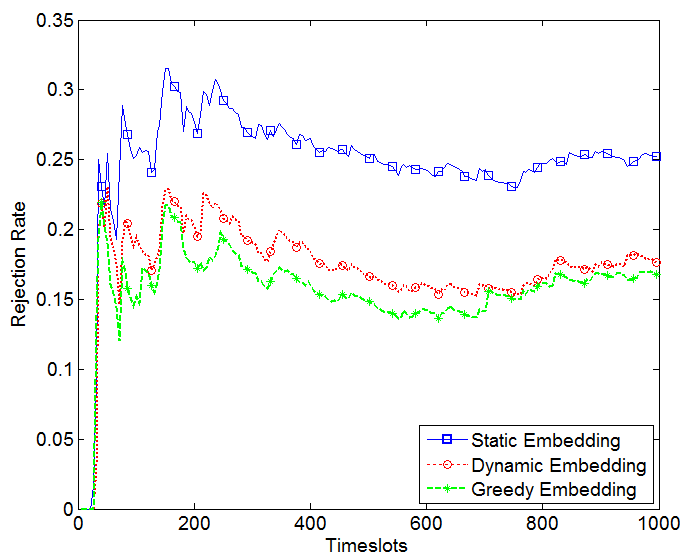}}\hfill
  \caption{A larger average request size leads to smaller revenues, and a higher number of rejected requests.}\label{fig:5}

\end{figure}

%

\subsubsection{Dynamic embedding achieves better performance than static embedding}In Figure \ref{fig:3} we compare the dynamic embedding algorithms and the static algorithm. Fig. \ref{fig:3}a shows the average revenue while Fig \ref{fig:3}b shows the average rejection rate of VNRs. After some initial fluctuation the graphs smooth off to the long-term values. It is clear that the dynamic embedding algorithms outperform the static algorithms in terms of revenue and rejection rate.

It is interesting to note that the rejection rate for the dynamic algorithms is 64\% lower on average than the static case but that the revenue is only 6.5\% better in the dynamic case. We can also note that the dynamic greedy algorithm achieves a slight increase in performance compared to the dynamic embedding. This shows that the use of dynamic algorithms benefits the virtual network operators as the rejection rate decreases significantly, but that for the infrastructure owner the additional revenue obtained is quite small.



\begin{figure}[!t]
\centering
   \includegraphics[width=.46\textwidth]{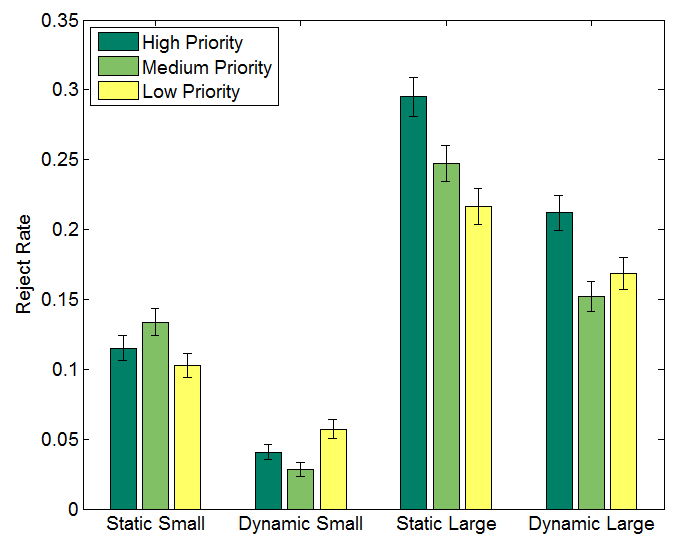}\hfill
  \caption{{Detailed rejection rates for the small and large scenarios show that the high priority rejection rate is much higher in the large scenario, relative to the other rejection rates.}}\label{fig:6}
\end{figure}

\subsubsection{Accuracy of resource request prediction}
Until now we have assumed that the virtual network operators making the VNRs know exactly what resources are required for each VNR. This may not be the case in practise. The accuracy of the predictions made by the VNOs influences the performance of the embedding algorithms. If the VNOs can make very accurate predictions then it is possible to have many requests with small resources requirements. However if it is not possible to predict accurately, it may be necessary to over-provision resources and make large requests.

We examine the case when the average resource requirements of the VNRs are higher, but the total number of resources requested remains the same as before. We set the resource requirement for each dimension to $U(2, 5)$ and reduce $\lambda$ to $1$. From Figure \ref{fig:5} we observe that the rejection rate is 1.7 times higher on average than the scenario where the requests were smaller but more numerous. The revenue is 13\% lower on average.  Fig \ref{fig:6} shows the mean reject rate and the standard error for the various priority levels and scenarios. From this figure, we can observe that a large number of high priority requests are rejected in the scenario where the requests are large. Since the high priority requests are the most valuable, this has the most impact on the revenue. We can conclude that when the VNOs are not able to make accurate predictions and the resource requests are large, this has a negative impact on both the rejection rate and the revenue.

\subsubsection{Priority costs}
The total revenue that can be achieved is influenced by the relative cost assigned to each priority level. However, the cost of each priority level also affects the number of VNRs received at that priority level. In the absence of a real market for wireless virtualization the optimum relative cost of each priority level cannot be known. In our model, since the demand for each priority level is uniform, the values assigned to $p_s$ do not influence the total revenue.



\section{Conclusion}
\label{sect:conclusion}
Many challenges still remain for wireless network virtualization. In this paper we presented a dynamic embedding method and formulated the embedding problem with the additional requirement of priority levels. We showed that better performance can be achieved using dynamic approaches.
Other challenges include solutions for heterogeneous networks, and the need for easy and implementable approaches. In future work, we hope to focus on a realistic pricing model and explore the architecture needed to make wireless network virtualization a reality.

\section*{Acknowledgment}

This material is based upon works supported by the Science Foundation
Ireland under Grants No. 10/CE/I1853 and 10/IN.1/I3007.

\bibliographystyle{IEEEtran}

\bibliography{Dynamic_embedding}

\end{document}